# Technical report: Kidney tumor segmentation using a 2D U-Net followed by a statistical post-processing filter

Iwan Paolucci

ARTORG Center for Biomedical Engineering Research, University of Bern,
Bern, Switzerland
iwan.paolucci@artorg.unibe.ch
http://www.artorg.unibe.ch

**Abstract.** Each year, there are about 400'000 new cases of kidney cancer worldwide causing around 175'000 deaths. For clinical decision making it is important to understand the morphometry of the tumor, which involves the time-consuming task of delineating tumor and kidney in 3D CT images. Automatic segmentation could be an important tool for clinicians and researchers to also study the correlations between tumor morphometry and clinical outcomes. We present a segmentation method which combines the popular U-Net convolutional neural network architecture with post-processing based on statistical constraints of the available training data. The full implementation, based on PyTorch, and the trained weights can be found on GitHub http://github.com/ipa/kits2019.

## 1 Introduction

Each year, there are about 400'000 new cases of kidney cancer worldwide causing around 175'000 deaths. Historically, radical nephrectomy which involves removal of the tumor and the kidney was considered gold standard for curative care. Recently, nephron sparing resection (partial nephrectomy) approaches are becoming more popular, as they result in better quality of life for the patient [1].

There are several scoring systems, which have a high predictive power towards surgical and clinical outcomes, even though they use rather simple characteristics of the tumor and the kidney derived from imaging. Therefore, it is assumed that more complex features have an even higher predictive power. Furthermore, automating the extraction of these features would be a clinically useful instrument for oncologists to decide on an appropriate surgical approach.

Recently, and especially since larger datasets are available, convolutional neural networks (CNN) have become the method of choice for image processing problems such as semantic segmentation. In this work, we present and evaluate a segmentation method based on the U-Net architecture, which is a CNN specifically designed for biomedical image segmentation.

## 2 Material and Methods

For this segmentation task, a two stage approach consisting of a segmentation step based on a U-Net and a post-processing step consisting of filtering and anatomical/statistical constraints is proposed Fig 1.

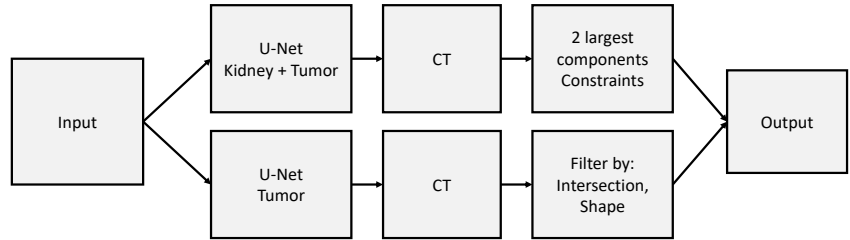

**Fig 1:** Concept of the whole image-processing pipeline

For training the dataset form the KiTS19 challenge was used [2]. This dataset is composed of 210 CT images of patients with kidney tumors with the corresponding segmentation masks for the kidney and the tumors. For efficient data loading, the slices of the CT images and the corresponding segmentation masks were extracted and stored separately as 2D images. The training data was randomly split into a training set (90%) and a test set (10%) on a case basis.

### 2.1 U-Net

The U-Net is a fully convolutional neural network (Fig 2), which was specifically designed for biomedical image-segmentation tasks [3]. In this work, we use this model for the initial segmentation of the kidneys and tumors. However, we applied some changes which are described in the following sections.

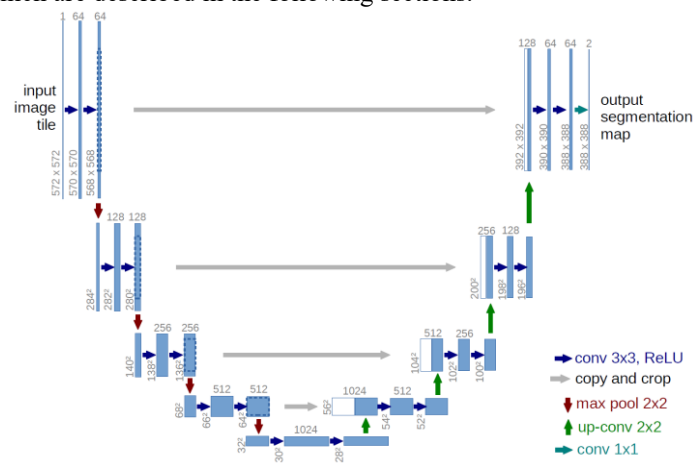

**Fig 2:** The original U-Net architecture consisting of an encoder and a decoder part with skip connections was used for this project [3]

Firstly, we used a single channel output with a sigmoid activation in the last layer, compared to a two-channel output with a soft-max activation. While this model is adaptable to multi-class problems, we trained two separate models – one for kidney + tumor and one for tumors only.

### 2.1.1 Training

In contrast to the original implementation we didn't use weights for the loss function but used a slightly adjusted loss function. The loss function (equation 1) is composed of a combination of the binary cross entropy (BCE) and the DICE loss function [4] with a weighting factor $\lambda$ for the DICE and 1- $\lambda$ for the BCE:

$$loss = (1 - \lambda) * BCE(pred, target) + \lambda * DICE(pred, target)$$
(1)

In this work, $\lambda$ was set to 0.5 to give both components of the loss function equal weight. We used this loss function with the Adam optimizer [5] and a learning rate scheduler, which adapts the learning rate based on the progress of the loss function. The model was trained from scratch and weights were initialized with random number from a gaussian distribution and the bias terms were set to 0. Both models were trained for 50 epochs, but both converged between 40 – 45 epochs and early stopped.

Firstly, we trained a 2D U-Net on the axial slices with input size 256 x 256, where the images were resized using nearest neighbor interpolation. The model was trained only on positive samples, where either kidney or tumor are visible. Additionally, the tumor labels were converted into kidney labels. Therefore, the model was trained to segment kidney + tumor.

Secondly, we trained another 2D U-Net on the lower half of the axial slices with input size 128 x 256, where the images were resized using nearest neighbor interpolation. The model was trained on positive tumor samples, where the tumor is visible. In contrast to the kidney model, the kidney labels where set to background and the tumor labels to foreground. Therefore, the model was trained to specifically segment the tumor.

Training on positive samples only causes the network to produce random output on negative samples at inference time. In some cases, even with a very strong activation in the last layer, causing false positives. However, we found that these can easily be filtered in a post-processing step.

During inference, each slice is passed through both models, and the results are then merged into a volume of the same size as the input volume, before passing the result to post-processing.

## 2.2 Post-processing

During the post-processing phase, holes in the segmentation are filled using morphological operators and false positive from the U-Net prediction are removed. After that, the two segmentation masks are merged into one volume.

For the kidney labels, a connected components labelling is applied to identify the two largest structures in the volume. With this simple anatomical constraint, we can eliminate the false positives that were caused by random activations on negative slices.

For the tumor labels, a statistical approach is used to remove segmentations, that do not fall in the shape distribution of kidney tumors. This statistical model is based on the ratio between the two largest eigenvalues of the tumor voxels. This ratio was first computed on all tumors available in the segmentations. The range of acceptable ratios was computed as the range between the 5$^{th}$ and 95$^{th}$ percentile of the ratio distribution. Specifically, this range was found to be between 1.07 and 2.8. Furthermore, a constraint that the tumor labels need to overlap with the kidney labels was applied. This is possible, because the first model segments kidney+tumor voxels. We chose a threshold of 95% overlap of the tumor with the kidney+tumor label, to accommodate for cases where the tumor segmentations are more accurate from the tumor model.

## 3 Experiments and Results

For evaluation of the segmentation method, the kidney and tumor DICE score of the segmentation on the test set were computed. In contrast to the training phase, where only positive samples were used, the whole CT volume was used and segmented. The test set consisted of 20 out of 210 datasets.

On this test set, the U-Net segmentation achieved a median DICE score of 0.91 and 0.49 for kidney+tumor and tumor respectively. After post-processing the mean DICE score was 0.97 and 0.61 for kidney+tumor and tumor respectively Fig 3.

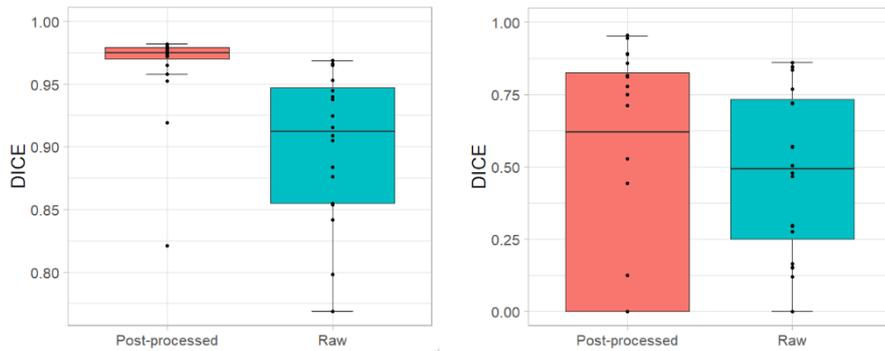

**Fig 3:** Results of the kidney + tumor segmentation (left) and the tumor segmentation (right) before (blue) and after post-processing (red)

The segmentation artifacts and the results after post-processing can be clearly seen in Fig 4. The post-processing step successfully removes false positives from the segmentation results of the kidney in almost all cases. However, it does remove some true positives from the tumor labels, even causing that in some cases all tumor detections are filtered out.

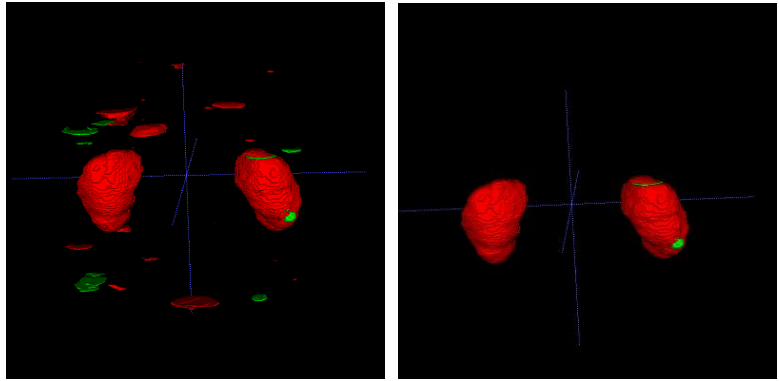

**Fig 4:** 3D segmentation result before (left) and after filtering of segmentation artifacts from random activations (right)

## 4 Conclusion

In this work, we presented how a simple CNN network followed by a post-processing step which purely applies anatomical constraints, can achieving competitive results. However, the algorithm failed in some cases of abnormal anatomy or with very large tumors. However, this is a small fraction of all cases which would require a radiologist' input when used in clinics.